\documentclass{optica-article}
\journal{oe}
\articletype{Research Article}
\begin{document}
\title{Quantum squeezing induced quantum entanglement and EPR steering in coupled optomechanical system}
\author{Shao-Xiong Wu,\authormark{1,2,*}, Cheng-Hua Bai,\authormark{1} Gang Li,\authormark{2,3} Chang-shui Yu,\authormark{4,\dag}, and Tiancai Zhang\authormark{2,3,\ddag}}
\address{\authormark{1} School of Semiconductor and Physics, North University of China, Taiyuan 030051, China\\
\authormark{2} State Key Laboratory of Quantum Optics and Quantum Optics Devices, and Institute of Opto-Electronics, Shanxi University, Taiyuan 030006, China\\
\authormark{3} Collaborative Innovation Center of Extreme Optics, Shanxi University, Taiyuan 030006, China\\
\authormark{4} School of Physics, Dalian University of Technology, Dalian 116024, China}
\email{\authormark{*}sxwu@nuc.edu.cn}
\email{\authormark{\dag}ycs@dlut.edu.cn}
\email{\authormark{\ddag}tczhang@sxu.edu.cn}

\begin{abstract}
We propose a theoretical project in which quantum squeezing induces quantum entanglement and Einstein-Podolsky-Rosen steering in a coupled whispering-gallery-mode optomechanical system. Through pumping the $\chi^{(2)}$-nonlinear resonator with the phase matching condition, the generated squeezed resonator mode and the mechanical mode of the optomechanical resonator can generate strong quantum entanglement and EPR steering, where the squeezing of the nonlinear resonator plays the vital role. The transitions from zero entanglement to strong entanglement and one-way steering to two-way steering can be realized by adjusting the system parameters appropriately. The photon-photon entanglement and steering between the two resonators can also be obtained by deducing the amplitude of the driving laser. Our project does not need an extraordinarily squeezed field, and it is convenient to manipulate and provides a novel and flexible avenue for diverse applications in quantum technology dependent on both optomechanical and photon-photon entanglement and steering.
\end{abstract}

\section{Introduction}
Quantum entanglement is not only one of the main striking features that distinguishes quantum physics from the classical world but also a fundamental resource in quantum information technology. In particular, the entanglement between microscopic and macroscopic matter is of significance to the basic theory of quantum mechanics both theoretically and experimentally. Due to the connection between microscopic matter (such as photons) and macroscopic matter (such as mechanical oscillators), the optomechanical system is a promising, flexible, and sensitive experimental platform to test and take advantage of the non-classical effects of quantum systems at the single-photon level \cite{am14}. The high quality optomechanical microcavities can be manufactured with the development of experimental technology, especially the micro/nano fabrication technique. How to realize strong and steady-state continuous entanglement is one of the essential issues in optomechanics research. After pioneering works on entanglement between a movable mirror and cavity field \cite{vd07,pm07}, optomechanical entanglement has been widely studied theoretically and experimentally and has becoming a booming research field. Scholars have realized strong steady-state entanglement in various ways based on different systems, such as auxiliary-cavity-assisted optomechanical systems \cite{ldg21}, generating entanglement by gently modulating optomechanical systems \cite{ma09,wm16}, using continuous measurement and feedback control \cite{md23}, nonreciprocal optomechanical entanglement in spin microresonator \cite{jyf20}, noise-tolerant optomechanical entanglement via synthetic magnetism \cite{ldg22}, ultizing the dual-mode cooling effect \cite{lzq21} or dark resonance \cite{hx15}, modulating two driving fields \cite{hcs20}, experimental demonstration by atomic ensembles \cite{okcf18}, and tri-particle optomechanical entanglement \cite{gc08,lj18,hxz21}.

As another kind of quantum non-locality, Einstein-Podolsky-Rosen (EPR) steering was proposed in the ideological experiment of the nonlocal problem between two-body systems, which denotes how to control or manipulate quantum entanglement \cite{ea35,se35,ur20}. In recent decades, the quantitative measurement of EPR steering has been mathematically defined \cite{rmd09,ki15}, where EPR steering is stronger than quantum entanglement but weaker than Bell nonlocality under the hierarchy of quantum nonlocality. Due to the intrinsic asymmetry, EPR steering may play a more fundamental role than quantum entanglement in nontrusted quantum communication. The properties of EPR steering and its applications \cite{hq15,fm21,try22,nrv20,dx17,ml23,ly23} in quantum technology have been widely reported. In optomechanical systems, EPR steering has also attracted attention, such as asymmetric EPR steering via a single dissipation pathway \cite{kd22}, enhancing EPR steering between two mechanical modes \cite{gq23}, manipulating asymmetric steering via interference effects \cite{zs19}, and EPR steering between distant macroscopic systems \cite{th21}. At the same time, the realization of quantum entanglement or EPR steering is often accompanied by squeezing or cooling of the mechanical mode, and realization squeezing beyond 3dB or ground state cooling of the mechanical oscillator have also been extensively reported \cite{lyc15,bch19,ljq11,zr19,wcw23,xb22,hj22,ags16,zw22pla,vp20,yj22}.

The squeezed state is a valuable resource in quantum optics and has many applications in various quantum processes. Under the phase matching conditions, the spontaneous parametric down conversion process can occur by employing $\chi^{(2)}$-nonlinear matter, and a squeezed process can be realized in the Bogoliubov transformation picture without an external squeezed field. In an optomechanical system, the nonlinear interaction between the squeezed cavity mode and mechanical mode can lead to a strong single-photon interaction \cite{lxy15}. A similar mechanism can be used to enhance the atom-field cooperativity parameter \cite{qw18}, strengthen the coherent dipole coupling between two atoms \cite{wy19}, explore the quantum speed limit in the squeezed optical cavity mode \cite{myj23}, and realize optical nonreciprocity induced by squeezing in a whispering-gallery-mode (WGM) cavity \cite{tl22}. As a kind of common optical cavity, the WGM cavity has been applied to demonstrate the non-Hermitian physics \cite{pb14,cl14}, the optomechanically-induced transparency phenomenon \cite{jh15}, the phase-controlled nonreciprocal routing \cite{sz23}, the phonon laser enhancement via optomechanics interaction \cite{lc23}, the nonreciprocal photon blockade in squeezed mode \cite{wdy23}, and the effect of optomechanical entanglement by exceptional point \cite{lz23}.

Inspired and encouraged by the above works, we systematically investigate the quantum entanglement and EPR steering induced by quantum squeezing in a coupled WGM optomechanical system in this paper. First, we focus on the entanglement and steering between the mechanical mode and the indirect interacted optical mode. Strong steady-state entanglement and steering can be achieved by properly optimizing the squeezing parameter, where the squeezing of the $\chi^{(2)}$-nonlinear resonator plays the central role in the generating quantum nonlocality. Second, we consider the photon-photon entanglement and steering between the coupled resonators by changing the amplitude of the driving laser. Unlike the quantum entanglement and EPR steering in cavity magnomechanical system via a squeezed vacuum field, the enhancement of quantum entanglement and EPR steering does not need an external squeezed field \cite{zw22,wsx23} or the method of reservoir engineering \cite{wyd13,lj15}.

The structure of this paper is as follows. In Sec. \ref{sec2}, we introduce the model of this project and derive the effective Hamiltonian by utilizing the quantum Langevin equation. In Sec. \ref{sec3}, we investigate the entanglement and steering between the $\chi^{(2)}$-nonlinear resonator and the mechanical oscillator, where the squeezing parameter plays an indispensable role in the generation of strong steady-state entanglement and steering, and discuss how to realize the transition from zero entanglement to strong entanglement and one-way steering to two-way steering. In Sec. \ref{sec4}, we study the entanglement and steering between the two resonator and show the transformation between steerable modes by changing the decay rate of resonators.

\begin{figure}[t]
\centering
\includegraphics[width=0.7\columnwidth]{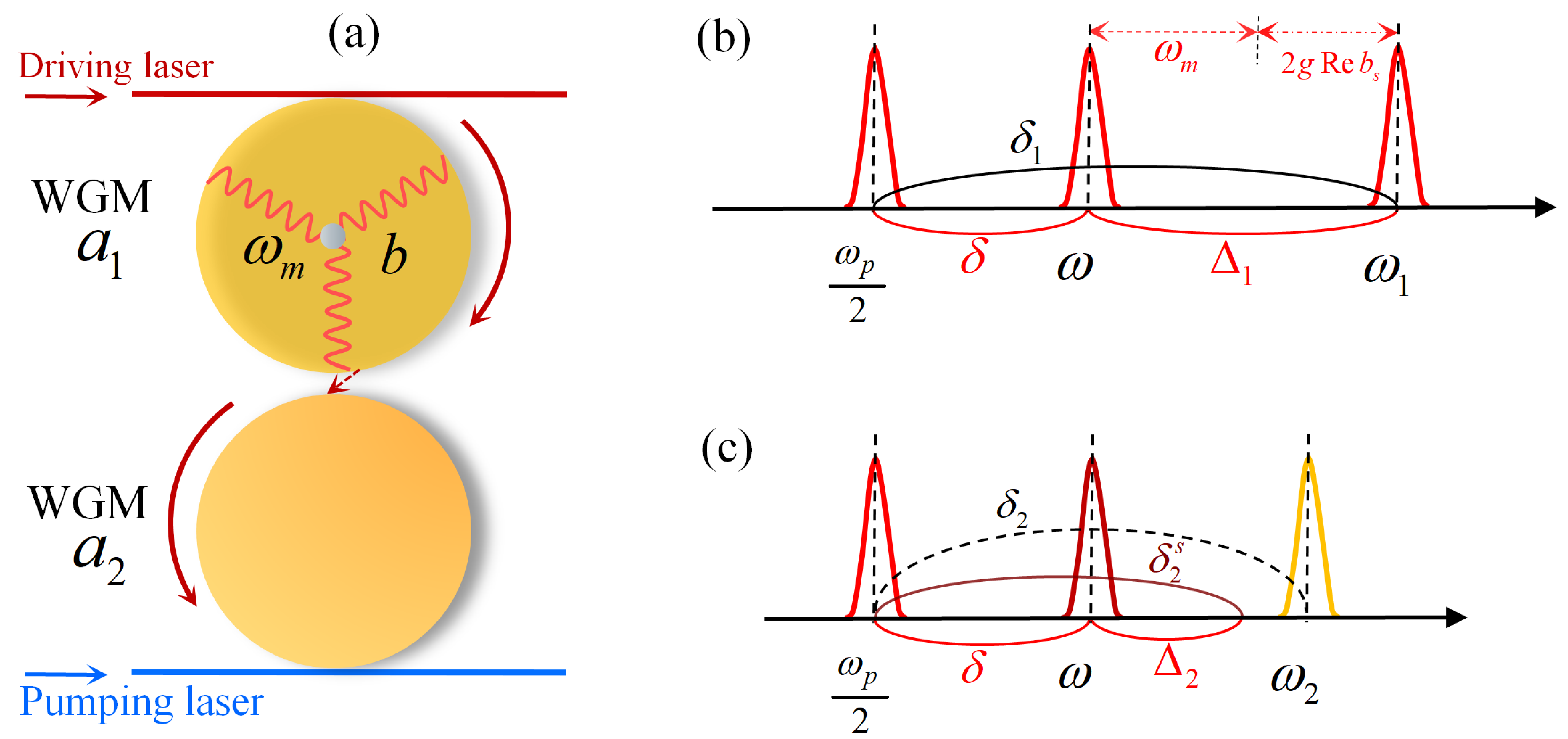}
\caption{(a) Sketched diagrams of the project. It consists of two coupled WGM resonators: the optomechanical resonator is driven by a coherent laser, and a $\chi^{(2)}$-nonlinear resonator is pumped under phase matching conditions. (b) The frequency relationship of resonator $a_1$. (c) The frequency relationship of resonator $a_2$. In the squeezing picture, a symbol $s$ is marked in the superscript.}
\label{fig1}
\end{figure}

\section{The model and linearized effective Hamiltonian}\label{sec2}
The coupled WGM optomechanical system model is depicted in Fig. \ref{fig1}(a). The Hamiltonian (in units of $\hbar$) of the whole system can be given as
\begin{align}
H_0=&\omega_1a_1^{\dagger}a_1+\omega_2a_2^{\dagger}a_2+\omega_mb^{\dagger}b-g a_1^{\dagger}a_1(b+b^{\dagger})+J(a_1^{\dagger}a_2+a_1a_2^{\dagger})\notag\\
&+iE(a_1^{\dagger}e^{-i\omega t}-a_1e^{i\omega t})+\frac{\Omega_p}{2}(e^{-i\theta}a_2^{\dagger 2}e^{-i\omega_pt}+e^{i\theta}a_2^2e^{i\omega_pt}),
\label{eqH0}
\end{align}
where $a_i$ and $\omega_i$ ($i=1,2$) express the annihilation operator and frequency of the $i$-th resonator's optical mode, and $b$ and $\omega_m$ denote the annihilation operator and frequency of the mechanical mode. The optomechanical resonator $a_1$ is driven by a coherent laser with amplitude $E$ and frequency $\omega$, and the single-photon coupling strength between the mechanical mode $b$ and optical mode $a_1$ is $g$. The coupling strength between the two optical resonators is $J$. The resonator $a_2$ is made up of high-quality materials with $\chi^{(2)}$ nonlinearity, e.g., lithium niobate or aluminum nitride \cite{jh15,tl22}, and pumped by a coherent laser with frequency $\omega_p$, which can generate squeezing interaction with strength $\Omega_p$ under phase-matching conditions through the parametric down-conversion process. Due to the different kinds of materials, detuning or phase matching conditions between resonators $a_1$ and $a_2$, the cavity mode of resonator $a_1$ is not squeezed.

In the frame rotating under frequency $\omega_p/2$, i.e., $U_0^{\dagger}H_0U_0-iU_0^{\dagger}\dot{U}_0$ with $U_0=\exp[-i\frac{\omega_p}{2}t(a_1^{\dagger}a_1+a_2^{\dagger}a_2)]$, the system Hamiltonian (\ref{eqH0}) can be rewritten as
\begin{align}
H_r=&\delta_1a_1^{\dagger}a_1+\delta_2a_2^{\dagger}a_2+\omega_mb^{\dagger}b-ga_1^{\dagger}a_1(b+b^{\dagger})+J(a_1^{\dagger}a_2+a_1a_2^{\dagger})\notag\\
&+iE(a_1^{\dagger}e^{-i\delta t}-a_1e^{i\delta t})+\frac{\Omega_p}{2}(e^{-i\theta}a_2^{\dagger 2}+e^{i\theta}a_2^2),
\label{eqHr}
\end{align}
where the detuning parameters are $\delta_1=\omega_1-\omega_p/2$, $\delta_2=\omega_2-\omega_p/2$, and $\delta=\omega-\omega_p/2$. Resonator $a_2$ can be transformed to the squeezing picture by applying the Bogoliubov transformation $a_2^s=S(r)a_2S^{\dagger}(r)=\cosh ra_2+e^{-i\theta}\sinh ra_2^{\dagger}$, where $S(r)=\exp[re^{i\theta}(a_2^2-a_2^{\dagger2})]$ is the single-mode squeezing operator and $r=\frac{1}{2}\text{arctanh}\beta$ is the squeezing parameter with a tunable coefficient $\beta=\Omega_p/\delta_2$. According to the inverse hyperbolic tangent function's domain, the amplitude $\Omega_p$  value should be smaller than the detuning of resonator $a_2$. In the squeezing picture, the rotating-wave approximation (RWA) can be employed under the weak coupling condition, i.e. $\delta_1+\delta_2^s\gg\sinh rJ$, and Hamiltonian (\ref{eqHr}) can be simplified as
\begin{align}
H_{\text{rwa}}=&\delta_1a_1^{\dagger}a_1+\delta_2^sa_2^{s\dagger}a_2^s+\omega_mb^{\dagger}b-ga_1^{\dagger}a_1(b+b^{\dagger})\notag\\
&+J^s(a_1^{\dagger}a_2^s+a_1a_2^{s\dagger})+iE(a_1^{\dagger}e^{-i\delta t}-a_1e^{i\delta t}),
\label{eqHrwa}
\end{align}
where $\delta_2^s=\delta_2\sqrt{1-\beta^2}$ is the detuning of resonator $a_2$ in the squeezing picture, and $J^s=\cosh rJ$ is the effective coupling strength between the two resonators.

Continuing to apply frame rotation with detuning $\delta$, Hamiltonian (\ref{eqHrwa}) can be described by
\begin{align}
H=&\Delta_1a_1^{\dagger}a_1+\Delta_2a_2^{s\dagger}a_2^s+\omega_mb^{\dagger}b-ga_1^{\dagger}a_1(b+b^{\dagger})\notag\\
&+J^s(a_1^{\dagger}a_2^s+a_1a_2^{s\dagger})+iE(a_1^{\dagger}-a_1)\label{eqH}
\end{align}
with the detuning parameters $\Delta_1=\delta_1-\delta$ and $\Delta_2=\delta_2^s-\delta$. The frequency relationship between the driving and pumping lasers and the resonators is shown vividly in Fig. \ref{fig1}(b) and (c). For simplicity and without loss of generality, the red detuning between the resonator $a_1$ and the driving laser, i.e., $\Delta_1=\omega_1-\omega$, is assumed to match the sum of the mechanical oscillator frequency $\omega_m$ and the displacement of detuning $2g\text{Re}b_s$ (see Eq. (\ref{eqa1sbs})) induced by the optomechanical interaction. In the following, we will investigate the dynamical behavior of the whole optomechanical system, and the symbol $a_2^s$ will be shorted as $a_2$ without ambiguity. In Fig. \ref{fig1}(c), $\delta_2$ expresses the detuning between the resonator $a_2$ and the pumping laser without the squeezing picture, expressed by the black dashed line. In contrast, the corresponding detuning in the squeezing picture is $\delta_2^s$, denoted by the crimson line.

Taking into account the dissipation and the corresponding environmental noise of the resonators $a_1$, $a_2$ and the mechanical oscillator $b$, the dynamics of the optomechanical system can be described by the nonlinear quantum Langevin equations \cite{gcw00,smo97}:
\begin{align}
\dot{a}_1=&(-i\Delta_1-\frac{\kappa_1}{2})a_1-iJ^sa_2+iga_1(b+b^{\dagger})+E+\sqrt{\kappa_1}a_1^{\text{in}},\notag\\
\dot{a}_2=&(-i\Delta_2-\frac{\kappa_2}{2})a_2-iJ^sa_1+\sqrt{\kappa_2}a_2^{\text{in}},\notag\\
\dot{b}=&(-i\omega_m-\frac{\gamma_m}{2})b+iga_1^{\dagger}a_1+\sqrt{\gamma_m}b^{\text{in}},
\label{eqqle1}
\end{align}
where $\kappa_i$ and $\gamma_m$ are the decay rates of resonator $a_i$ and the mechanical oscillator, respectively, and $a_i^{\text{in}}$ and $b^{\text{in}}$ are the corresponding zero-mean environment noise operators. Notice that the decay rate $\kappa_2$ will not be changed in the squeezing picture.

The quantum Langevin equations in Eq. (\ref{eqqle1}) can be solved by the standard linearization process under a strong external coherent driving laser, and the operators of resonator $a_i$ and mechanical oscillator $b$ can be rewritten as the sum of steady-state mean values and quantum fluctuation operators, i.e., $o=o_s+\delta o$ $(o=a_i,b)$. The differential equations for the steady-state mean values are
\begin{align}
\dot{a}_{1s}&=(-i\Delta_1'-\frac{\kappa_1}{2})a_{1s}-iJ^sa_2+E,\notag\\
\dot{a}_{2s}&=(-i\Delta_2-\frac{\kappa_2}{2})a_{2s}-iJ^sa_{1s},\notag\\
\dot{b}_s&=(-i\omega_m-\frac{\gamma_m}{2})b_s+ig|a_{1s}|^2,
\label{eqa1sa2sbs}
\end{align}
where the effective detuning $\Delta_1'=\Delta_1-2g\text{Re}b_s$. For the steady state, its operator mean value is not changed with time, and Eq. (\ref{eqa1sa2sbs}) can be solved as
\begin{align}
a_{1s}=\frac{E(i\Delta_2+\frac{\kappa_2}{2})}{J^{s2} +(i\Delta_1'+\frac{\kappa_1}{2})(i\Delta_2+\frac{\kappa_2}{2})}, b_s=\frac{ig|a_{1s}|^2}{i\omega_m+\frac{\gamma_m}{2}}.
\label{eqa1sbs}
\end{align}
By adjusting the phase of the driving laser, the value of $a_{1s}$ can be considered as a real number, and the following form determines the linearized quantum Langevin equations for the quantum fluctuation operators:
\begin{align}
\dot{\delta a_1}=&(-i\Delta_1'-\frac{\kappa_1}{2})\delta a_1-iJ^s\delta a_2+iga_{1s}(\delta b+\delta b^{\dagger})+\sqrt{\kappa_1}a_1^{\text{in}},\notag\\
\dot{\delta a_2}=&(-i\Delta_2-\frac{\kappa_2}{2})\delta a_2-iJ^s\delta a_1+\sqrt{\kappa_2}a_2^{\text{in}},\notag\\
\dot{\delta b}=&(-i\omega_m-\frac{\gamma_m}{2})\delta b+iga_{1s}(\delta a_1+\delta a_1^{\dagger})+\sqrt{\gamma_m}b^{\text{in}}.
\label{eqqle2}
\end{align}

According to Eq. (\ref{eqqle2}), the effective Hamiltonian of quantum fluctuation operators can be obtained as
\begin{align}
H_{\text{eff}}=&\Delta_1'\delta a_1^{\dagger}\delta a_1+\Delta_2\delta a_2^{\dagger}\delta a_2+J^s(\delta a_1\delta a_2^{\dagger}+\delta a_1^{\dagger}\delta a_2)\notag\\
&+\omega_m\delta b^{\dagger}\delta b-G(\delta a_1+\delta a_1^{\dagger})(\delta b^{\dagger}+\delta b)
\label{eqHeff}
\end{align}
with the effective optomechanical coupling strength $G=ga_{1s}$. For the weak coupling regime and employing the RWA approximation, the effective linearized Hamiltonian arrives at
\begin{align}
H_{\text{lin}}=&\Delta_1'\delta a_1^{\dagger}\delta a_1+\Delta_2\delta a_2^{\dagger}\delta a_2+J^s(\delta a_1\delta a_2^{\dagger}+\delta a_1^{\dagger}\delta a_2)\notag\\
&+\omega_{m}\delta b^{\dagger}\delta b-G(\delta a_1\delta b^{\dagger}+\delta a_1^{\dagger}\delta b).
\label{eqHlin}
\end{align}
The coefficients in Hamiltonian (\ref{eqHlin}), i.e., the effective detuning $\Delta_1'$, $\Delta_2$, the effective coupling strength $J^s$, and the effective optomechanical coupling strength $G$, are affected by the squeezing of the resonator $a_2$. For simplicity and without loss of generality, the value of the effective detuning $\Delta_1'$ is assumed to be the same as that of $\omega_m$ in what follows.

Because of the squeezing transformation of the resonator $a_2$, the nonzero noise correlation functions of operator $a_2^{\text{in}}$ will be turned into the squeezing picture and read $\langle a_2^{\text{in}\dagger}(t)a_2^{\text{in}}(\tau)\rangle=N\delta(t-\tau)$, $\langle a_2^{\text{in}}(t)a_2^{\text{in}\dagger}(\tau)\rangle=(N+1)\delta(t-\tau)$, $\langle a_2^{\text{in}}(t)a_2^{\text{in}}(\tau)\rangle=M\delta(t-\tau)$, and $\langle a_2^{\text{in}\dagger}(t)a_2^{\text{in}\dagger}(\tau)\rangle=M^*\delta(t-\tau)$ with $N=\sinh^2r$ and $M=\cosh r\sinh r\exp(-i\theta)$ under the Markovian approximation \cite{gcw00,smo97}. The non-zero noise correlation functions of operators $a_1^{\text{in}}$ and $b^{\text{in}}$ meet: $\langle a_1^{\text{in}\dagger}(t)a_1^{\text{in}}(\tau)\rangle=0$, $\langle a_1^{\text{in}}(t)a_1^{\text{in}\dagger}(\tau)\rangle=\delta(t-\tau)$, $\langle b^{\text{in}\dagger}(t)b^{\text{in}}(\tau)\rangle=m\delta(t-\tau)$, and $\langle b^{\text{in}}(t)b^{\text{in}\dagger}(\tau)\rangle=(m+1)\delta(t-\tau)$, where $m=[\exp(\hbar\omega_m/k_BT)-1]^{-1}$ is the thermal mean phonon number of mechanical oscillator $b$ with temperature $T$.

To investigate the dynamics of the system, we will define the position and momentum quadrature fluctuation operators and noise operators as
\begin{align}
X_o=\frac{o+o^{\dagger}}{\sqrt{2}},& Y_o=\frac{o-o^{\dagger}}{i\sqrt{2}},\notag\\
X_{o^{\text{in}}}=\frac{o^{\text{in}}+o^{\text{in}\dagger}}{\sqrt{2}},& Y_{o^{\text{in}}}=\frac{o^{\text{in}}-o^{\text{in}\dagger}}{i\sqrt{2}},
\label{eqXY}
\end{align}
where the operators are $o=\delta a_1,\delta a_2,\delta b$ and $o^{\text{in}}=a_1^{\text{in}},a_2^{\text{in}},b^{\text{in}}$. The vector of the quadrature fluctuation operators is set as $\mathcal{R}(t)=[X_{\delta a_1}, Y_{\delta a_1}, X_{\delta a_2}, Y_{\delta a_2}, X_{\delta b},  Y_{\delta b}]^{T}$, and the corresponding vector of the quadrature noise operators is $\mathcal{N}(t)=[X_{a_1^{\text{in}}}, Y_{a_1^{\text{in}}}, X_{a_2^{\text{in}}}, Y_{a_2^{\text{in}}},  X_{b^{\text{in}}}, Y_{b^{\text{in}}}]^{T}$. Based on the linearized quantum Langevin equations (\ref{eqqle2}), the dynamics of the optomechanical system can be converted to a matrix form
\begin{align}
\frac{d\mathcal{R}(t)}{dt}=\mathcal{M}(t)\mathcal{R}(t)+\mathcal{N}(t),
\end{align}
where the drift matrix $\mathcal{M}(t)$ is a $6\times6$ matrix with the following form
\begin{align}
\mathcal{M}(t)=\left[\begin{array}{cccccc}
-\frac{\kappa_1}{2} & \Delta_1' & 0 & J^s & 0 & -G\\
-\Delta_{1}' & -\frac{\kappa_1}{2} & -J^s & 0 & G & 0\\
0 & J^s & -\frac{\kappa_2}{2} & \Delta_2 & 0 & 0\\
-J^s & 0 & -\Delta_2 & -\frac{\kappa_2}{2} & 0 & 0\\
0 & -G & 0 & 0 & -\frac{\gamma_m}{2} & \omega_m\\
G & 0 & 0 & 0 & -\omega_m & -\frac{\gamma_m}{2}
\end{array}\right].
\label{eqM}
\end{align}

Since the dynamical evolution of the quadrature fluctuation operators (\ref{eqXY}) is governed by the linearized Hamiltonian $H_{\text{lin}}$ (\ref{eqHlin}), the quadrature fluctuation operators can form a three-mode Gaussian state due to the intrinsic Gaussian nature of quantum noise. The Gaussian state can be characterized by a $6\times 6$ covariance matrix $\mathcal{V}(t)$ with component $\mathcal{V}_{kl}=\langle\mathcal{R}_k\mathcal{R}_l+\mathcal{R}_l\mathcal{R}_k\rangle/2$, and its dynamics are determined by the motion equation:
\begin{align}
\frac{d\mathcal{V}}{dt}=\mathcal{M}\mathcal{V}+\mathcal{V}\mathcal{M}^T+\mathcal{D},
\label{eqdV}
\end{align}
where $\mathcal{D}$ is the noise diffusion matrix with element $\mathcal{D}_{kl}=\langle\mathcal{N}_k\mathcal{N}_l+\mathcal{N}_l\mathcal{N}_k\rangle/2$. According to the correlation functions, the noise diffusion matrix $\mathcal{D}$ has a direct sum form
\begin{align}
\mathcal{D}=\mathcal{D}_1\oplus\mathcal{D}_2\oplus\mathcal{D}_3
\label{eqD}
\end{align}
with subblock matrices
\begin{align}
\mathcal{D}_1&=\frac{\kappa_1}{2}\text{diag}[1,1],\notag\\
\mathcal{D}_2&=\frac{\kappa_2}{2}
\left[\begin{array}{cc}
2N+1+M+M^{*} & i(M^{*}-M)\\
i(M^{*}-M) & 2N+1-M-M^{*}
\end{array}\right],\notag\\
\mathcal{D}_3&=\frac{\gamma_m}{2}\text{diag}[2m+1,2m+1].
\label{eqD1D2D3}
\end{align}

The reduced two-mode state among the three-mode Gaussian state can be described by a $4\times 4$ covariance matrix with block matrix form
\begin{align}
\mathcal{V}_{12}=\left[
                   \begin{array}{cc}
                     \mathcal{V}_1 & \mathcal{V}_c \\
                     \mathcal{V}_c^T & \mathcal{V}_2 \\
                   \end{array}
                 \right],
\end{align}
where the $2\times2$ subblock matrices $\mathcal{V}_1$, $\mathcal{V}_2$ and $\mathcal{V}_c$ correspond to the subblock covariance matrices of modes 1 and 2 and their correlated parts, respectively. The two-mode Gaussian entanglement can be measured by logarithmic Negativity \cite{vg02,ag04,pmb05}:
\begin{align}
E_N=\max[0,-\ln2\eta^-]
\end{align}
with the coefficients $\eta^-=\sqrt{\Sigma-\sqrt{\Sigma^2-4\det\mathcal{V}_{12}}}/\sqrt{2}$ and $\Sigma=\det \mathcal{V}_1+\det \mathcal{V}_2-2\det \mathcal{V}_c$. As another nonlocality measure, the EPR steering from mode 1 to 2, i.e., mode 1 can steer mode 2, is defined \cite{ki15} as
\begin{align}
\mathcal{G}_{1\rightarrow2}= \max\left[0,\frac{1}{2}\ln\frac{\det\mathcal{V}_1}{4\det\mathcal{V}_{12}}\right].
\end{align}
Similarly, the EPR steering from mode 2 to 1 is $\mathcal{G}_{2\rightarrow1}= \max[0,\frac{1}{2}\ln\frac{\det\mathcal{V}_2}{4\det\mathcal{V}_{12}}]$. In the following, we will investigate the influence of quantum squeezing of resonator $a_2$ on the optomechanical system's quantum entanglement and EPR steering.

\section{Entanglement and steering of resonator $a_2$ and mechanical oscillator $b$}\label{sec3}
\begin{figure}[t]
\centering
\includegraphics[width=1\columnwidth]{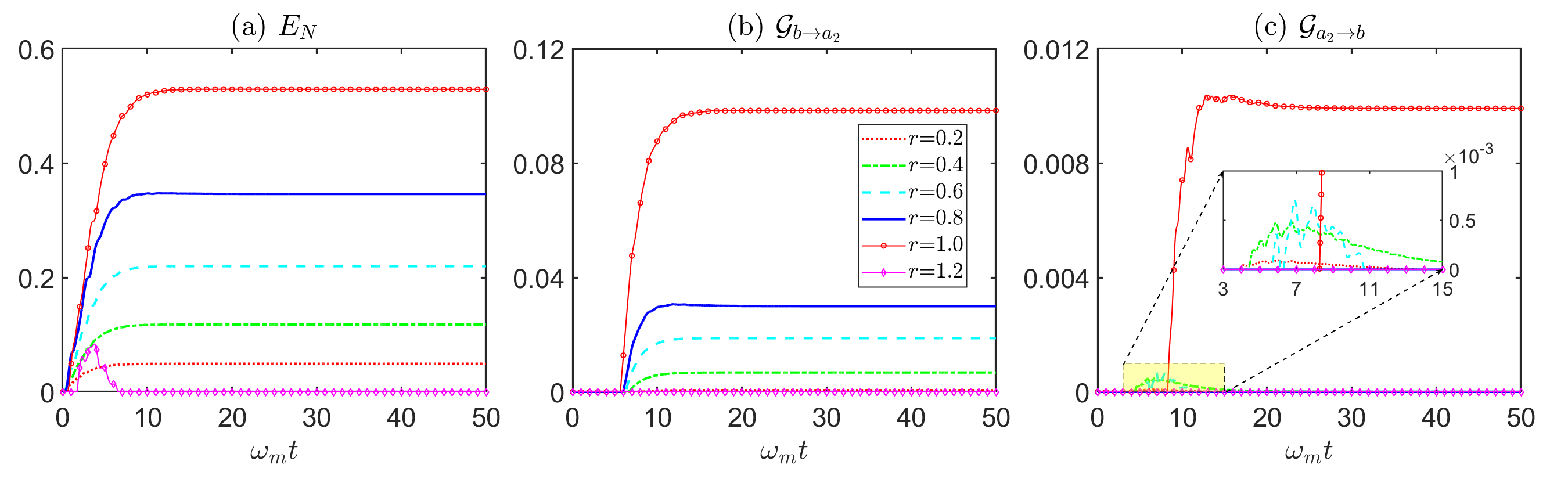}
\caption{The dynamics of entanglement $E_N$ (a), steering $\mathcal{G}_{b\rightarrow a_2}$ (b), and $\mathcal{G}_{a_2\rightarrow b}$ (c) as functions of time. The frequency of the mechanical oscillator is $\omega_m=2\pi\times 23.4\text{MHz}$, and the other parameters are set: $\kappa_1/\omega_m=0.6$, $\kappa_2=\kappa_1$, $\gamma_m/\omega_m=10^{-5}$, $J/\omega_m=1$, $g/\omega_m=8.5\times10^{-5}$, and $E/\omega_m=3.7\times10^5$.}
\label{fig2}
\end{figure}

In this section, we will first consider the optomechanical entanglement between the resonator $a_2$ and the mechanical oscillator $b$ based on the effective linear Hamiltonian, which is indirectly interacted through the beam-splitter interaction among the modes $\delta a_1-\delta b$ and $\delta a_1-\delta a_2$. In Fig. \ref{fig2}, we plot the dynamics of the quantum entanglement $E_N$ (a) and EPR steering $\mathcal{G}_{b\rightarrow a_2}$ (b), $\mathcal{G}_{a_2\rightarrow b}$ (c) between the resonator $a_2$ and mechanical oscillator $b$ under different squeezing parameters. Based on the current experimental technique, the system parameters can be chosen as: $\omega_m=2\pi\times 23.4\text{MHz}$, $\kappa_1/\omega_m=0.6$, $\kappa_2=\kappa_1$, $\gamma_m/\omega_m=10^{-5}$, $J/\omega_m=1$, $g/\omega_m=8.5\times10^{-5}$, and $E/\omega_m=3.7\times10^5$. The power of the 1550nm continuous wave tunable driving laser is $P_{\text{in}}=\omega E^2/\kappa\simeq6.9\text{mW}$, and the squeezing of the resonator mode $\delta a_2$ induced by the pumping laser is essential to generate quantum nonlocality in both quantum entanglement and EPR steering. When the pumping laser is absent, i.e. $r=0$, the entanglement and steering are zero. As the pumping strength $\Omega_p$ increases, the squeezing parameter $r$ will be enhanced. For $r=0.2$ (red dotted line), there exists steady entanglement; however, the steering  $\mathcal{G}_{b\rightarrow a_2}$ is zero, while there is transient but unstable steering $\mathcal{G}_{a_2\rightarrow b}$ in the dynamical evolution (shown in the inset of panel (c)). When the squeezing strength $r$ is continued to be raised, such as $r=0.4$ (green dotted-dashed line) or $0.6$ (cyan dashed line), there is steady entanglement and steering $\mathcal{G}_{b\rightarrow a_2}$, while the behavior of the steering $\mathcal{G}_{a_2\rightarrow b}$ is similar to that under the condition $r=0.2$ and the steady steering $\mathcal{G}_{a_2\rightarrow b}$ is zero. As $r$ increases to 0.8 (blue solid line), the value of steady entanglement $E_N$ and steering $\mathcal{G}_{b\rightarrow a_2}$ will increase with the squeezing parameter $r$; however, there is no steering $\mathcal{G}_{a_2\rightarrow b}$ in the whole dynamical evolution. When the squeezing parameter is $r=1$ (red circle line), both the value of entanglement $E_N$ and steering $\mathcal{G}_{b\rightarrow a_2}$ can be greatly enhanced; more importantly, steady steering $\mathcal{G}_{a_2\rightarrow b}$ exists. However, when the squeezing parameter becomes $r=1.2$ (magenta diamond line), there is no steady entanglement and steering. One can find that the squeezing of resonator $a_2$ plays a crucial role in the generating entanglement and steering between the resonator mode $\delta a_2$ and mechanical mode $\delta b$.

The physical mechanism interpretation behind the steady entanglement may be explained as follows. For the red-detuning driving, the Hamiltonian (\ref{eqHlin}) can be dealt with in principle as the form $\eta(\delta a_1\beta^++\delta a_1^{\dagger}\beta)$ with the Bogoliubov mode $\beta=\cosh\lambda\delta b+\sinh\lambda\delta a_2^{\dagger}$, where $\eta=\sqrt{G^2-J^{s2}}$ is the coupling strength between the resonator mode $\delta a_1$ and Bogoliubov mode $\beta$, and $\lambda=\text{arctanh}(J^s/G)$ can be considered as a squeezing coefficient of the two-mode squeezing operator $S(\lambda)=\exp[\lambda(\delta b\delta a_2-\delta b^{\dagger}\delta a_2^{\dagger})]$. According to the Duan-Simon criterion \cite{dlm00,sr00}, the modes $\delta b$ and $\delta a_2$ of the Bogoliubov mode $\beta$ are entangled, which is closely related to the squeezing coefficient $\lambda$. Due to the beam-splitter interaction, the Bogoliubov mode $\beta$ can be cooled by the resonator mode $\delta a_1$. Intuitively, the steady entanglement between the modes $\delta b$ and $\delta a_2$ is maximized when the Bogoliubov mode $\beta$ is cooled to the ground state through the resonator mode $\delta a_1$, and the squeezing coefficient $\lambda$ should be as large as possible. However, the coupling strength $\eta$ between the resonator mode $\delta a_1$ and Bogoliubov mode $\beta$  will decrease with increasing effective coupling strength $J^s$ between the resonators $a_1$ and $a_2$, which will reduce the cooling effect and limit the generation of entanglement. The coefficients $J^s$ and $G$ are related to the squeezing parameter $r$ induced by the pumping laser on the resonator $a_2$, and strong quantum entanglement and EPR steering can be obtained by balancing the comprehensive interaction and optimizing the squeezing parameter $r$, which is determined by the pumping laser.

\begin{figure}[h]
\centering
\includegraphics[width=0.75\columnwidth]{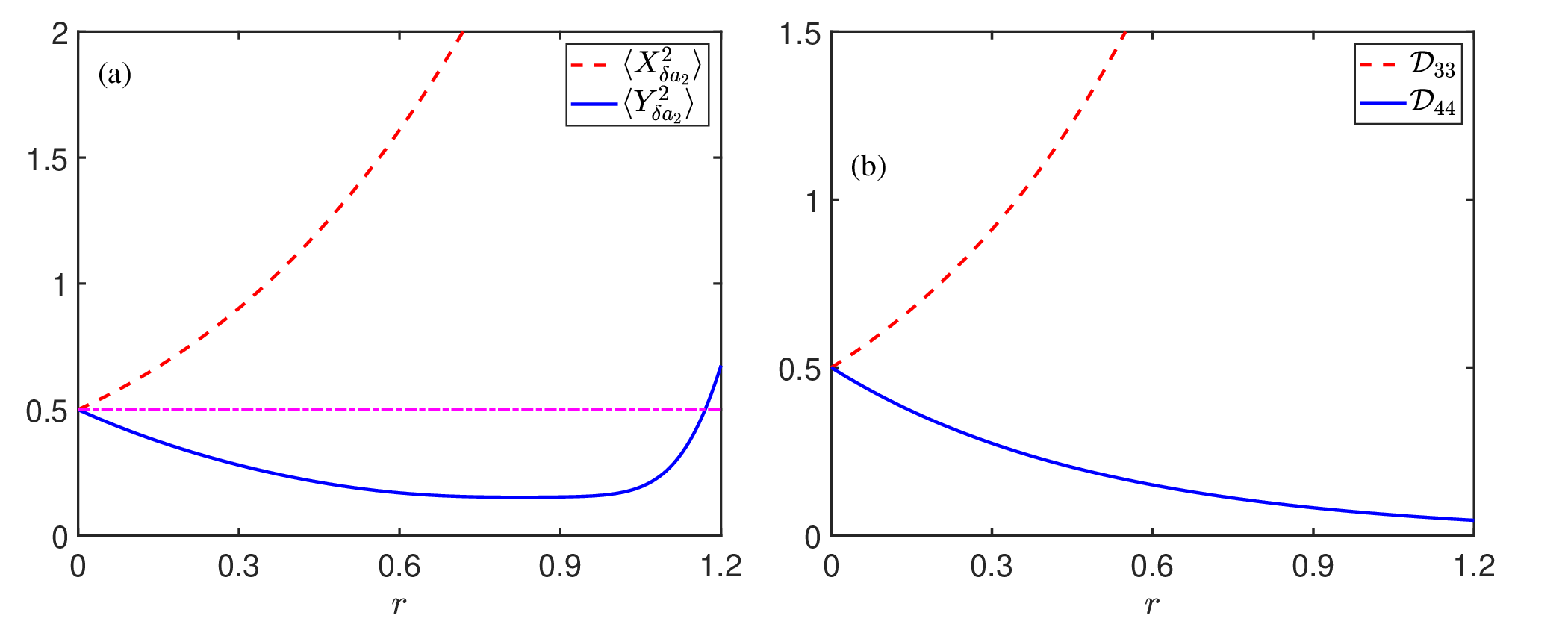}
\caption{(a) The variance of $\langle X_{\delta a_2}\rangle$ and $\langle Y_{\delta a_2}\rangle$ as a function of squeezing parameter $r$. (b) The variation (in units of $\kappa_1$) of the elements of noise matrix $\mathcal{D}_{33}$ and $\mathcal{D}_{44}$ along with the squeezing parameter $r$.}
\label{fig3}
\end{figure}

According to the generation mechanism of quantum entanglement and steering between the resonator mode $\delta a_2$ and mechanical mode $\delta b$, we plot the variation in the variances of position $\langle X_{\delta a_2}^2\rangle$ and moment $\langle Y_{\delta a_2}^2\rangle$ along with the squeezing parameter $r$ in Fig. \ref{fig3}(a). It is easy to find that the variance of position $\langle X_{\delta a_2}^2\rangle$ increases monotonically with increasing $r$. The variance of moment $\langle Y_{\delta a_2}^2\rangle$ decreases at first, reaches a minimum value of 0.152 (corresponding to $r\simeq0.84$) and maintains a small value until $r\simeq1$. Then, the value of $\langle Y_{\delta a_2}^2\rangle$ grows and returns back to 0.5 for $r\simeq1.18$, and the moment operator cannot be squeezed. One can also check that there is no squeezing of the position and the moment operator for the resonator mode $\delta a_1$ and mechanical mode $\delta b$. The generation of entanglement is accompanied by the occurrence of squeezing, so the moment operator squeezing of the resonator mode $\delta a_2$ induces entanglement with the mechanical mode $\delta b$ through the resonator mode $\delta a_1$ as an intermediate interaction. The squeezing is the necessary condition for generating entanglement in this project. The strength of entanglement is strongly related to the squeezing parameter $r$. To analyze the reason why the moment operator of mode $\delta a_2$ can be squeezed, we plot the dynamics of the elements of the noise subblock diffusion matrix $\mathcal{D}_{33}$ and $D_{44}$ in Fig. \ref{fig3}(b). Under a constant phase, for example, $\theta=0$, the behavior of $\mathcal{D}_{33}$ and $\mathcal{D}_{44}$ exhibit opposite tendencies with increasing $r$, and $\mathcal{D}_{33}$ acts like a heating bath, which will lead the variance of position $\langle X_{\delta a_2}^2\rangle$ to continue to increase, while $\mathcal{D}_{44}$ acts like a cooling bath and makes the value of $\langle Y_{\delta a_2}^2\rangle$ decrease. However, the variance of moment $\langle Y_{\delta a_2}^2\rangle$ is also influenced by $\mathcal{D}_{33}$ and shows behavior that decrease first and then increases as $r$ increases under the two opposite competing comprehensive effects, and the squeezing of the noise matrix is transferred to the mode $\delta a_2$, which contributes to the entanglement between the modes $\delta a_2$ and $\delta b$.

\begin{figure}[t]
\centering
\includegraphics[width=1.0\columnwidth]{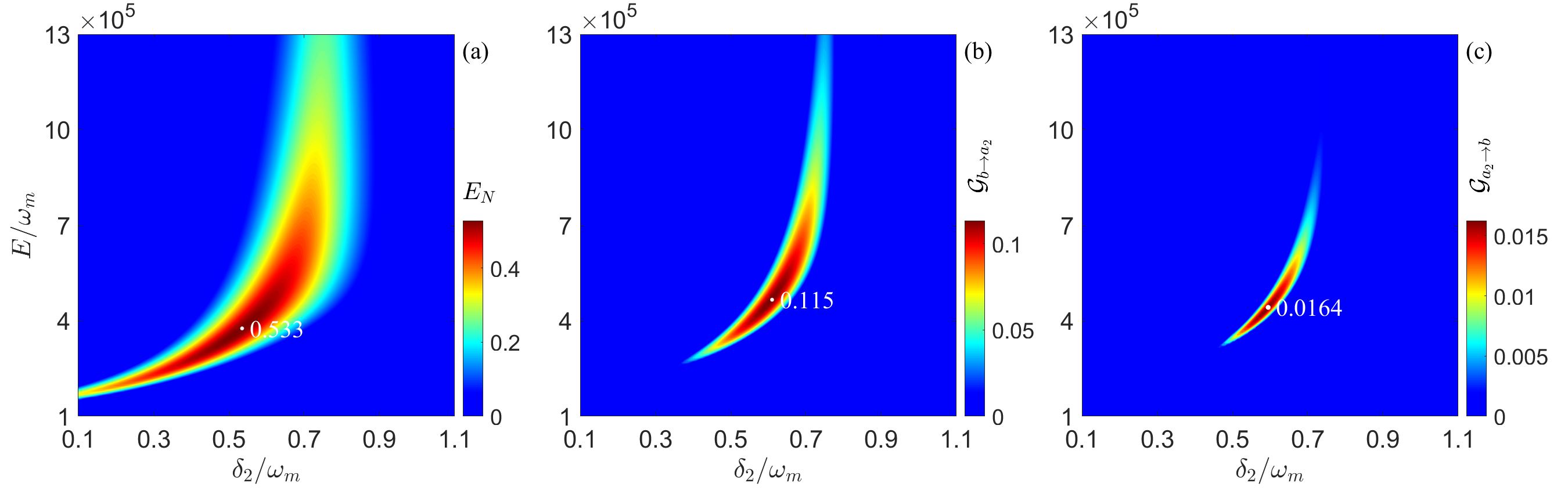}
\includegraphics[width=1.0\columnwidth]{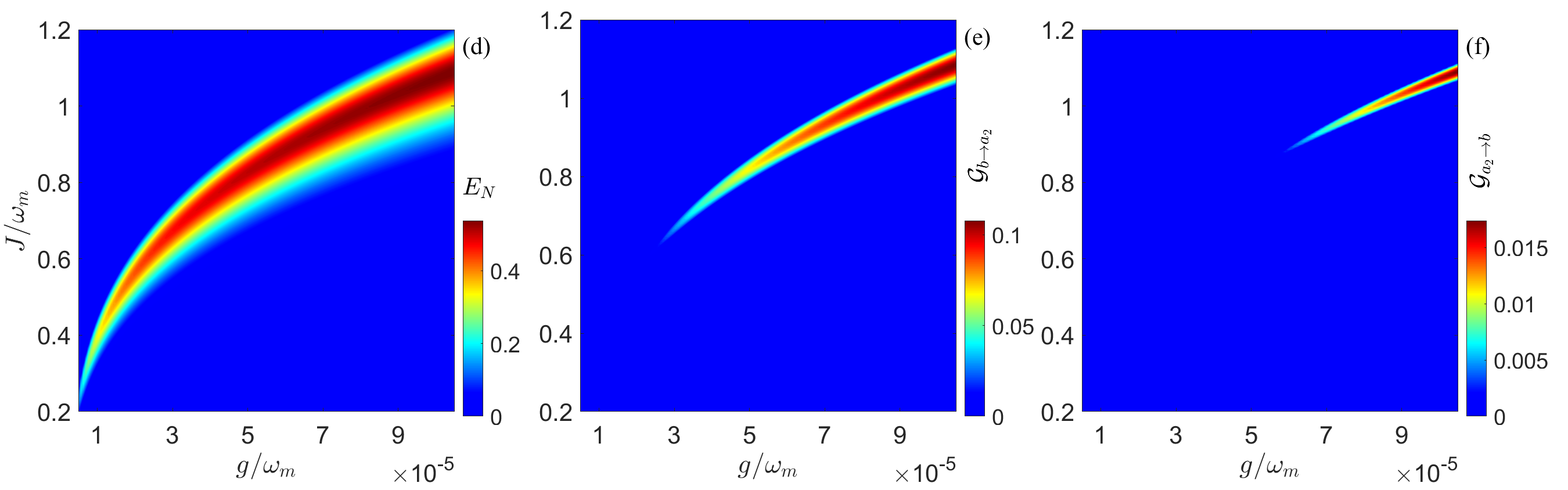}
\caption{The dynamics of entanglement $E_N$ (a), steering $\mathcal{G}_{b\rightarrow a_2}$ (b), and $\mathcal{G}_{a_2\rightarrow b}$ (c) as functions of the detuning $\delta_2$ and driving amplitude $E$. The variation in the corresponding nonlocality as functions of the single-photon coupling $g$ and resonators coupling $J$ are shown in (e)-(f). The system parameters are chosen as those in Fig. \ref{fig2}.}
\label{fig4}
\end{figure}

Since steady correlation is essential in quantum technology, we will focus on steady entanglement and steering in the following content. According to the above analysis, quantum entanglement and EPR steering can be generated and enhanced with adequate squeezing parameter $r$. In Fig. \ref{fig4}, we plot the variation of entanglement $E_N$ and steering $\mathcal{G}_{b\rightarrow a_2}$ and $\mathcal{G}_{a_2\rightarrow b}$ versus the detuning parameter $\delta_2$ and drive amplitude $E$ in panels (a)-(c). For comparison, the behaviour of entanglement and steering as functions of the single-photon coupling strength $g$ and coupling strength $J$ between the resonators are shown in panels (d)-(f). In all panels, the squeezing parameter is set as $r=1$, and the corresponding strength of the pumping laser is $\Omega_p/\omega_m=0.5$ with power $P_p=\frac{1}{16}\Omega_p^2\kappa_p\omega_p/\chi^2\simeq1.1\text{nW}$ under detuning $\delta_2/\omega_m=0.52$, $\delta/\omega_m=0.5$ as an example \cite{tl22}. The other parameters are the same as those in Fig. \ref{fig2}, and the Routh-Hurwitz criterion is applied throughout the article to ensure the system is in the stable regime \cite{dex87}.

In Figs. \ref{fig4}(a)-(c), strong steady-state entanglement and steering can be obtained in a broad regime by controlling the detuning $\delta_2$ between the pumping laser and resonator $a_2$ and amplitude $E$ of the driving laser of the resonator $a_1$. However, a stronger amplitude is not always better, and the maximum values, i.e., $E_N\simeq0.533$, $\mathcal{G}_{b\rightarrow a_2}\simeq0.115$, and $\mathcal{G}_{a_2\rightarrow b}\simeq0.0164$, are marked. In Hamiltonian (\ref{eqHlin}), the effective detuning $\Delta_2$ is related to the squeezing parameter $r$ and grows with the detuning $\delta_2$ homologically. In contrast, the effective coupling strength $G$ will be enhanced by increasing both $\Delta_2$ and $E$ and is majorly influenced by the driving amplitude $E$ due to the magnitude difference. With increasing $G$, the squeezing coefficient $\lambda$ in the Bogoliubov mode $\beta$ will decrease monotonically, and the entanglement will reduce seemingly. However, based on the physical mechanism analysis, increasing $G$ will strengthen the effective interaction $\eta$ between the resonator mode $\delta a_1$ and Bogoliubov mode $\beta$, which will lead to a stronger cooling effect on the Bogoliubov mode $\beta$. The result of the competitive and cooperative effects between these two opposite effects on entanglement is shown in panel (a). In essence, both EPR steering and quantum entanglement stem from the nonlocality of quantum mechanics. However, EPR steering is stronger than quantum entanglement under the hierarchy of quantum nonlocality, so the nonzero steering regions are smaller than those of quantum entanglement. Moreover, due to the intrinsic asymmetrical feature of EPR steering, one can also investigate one-way and two-way steering between modes $\delta a_2$ and $\delta b$ (see Fig. \ref{fig5} below).

\begin{figure}[b]
\centering
\includegraphics[width=0.9\columnwidth]{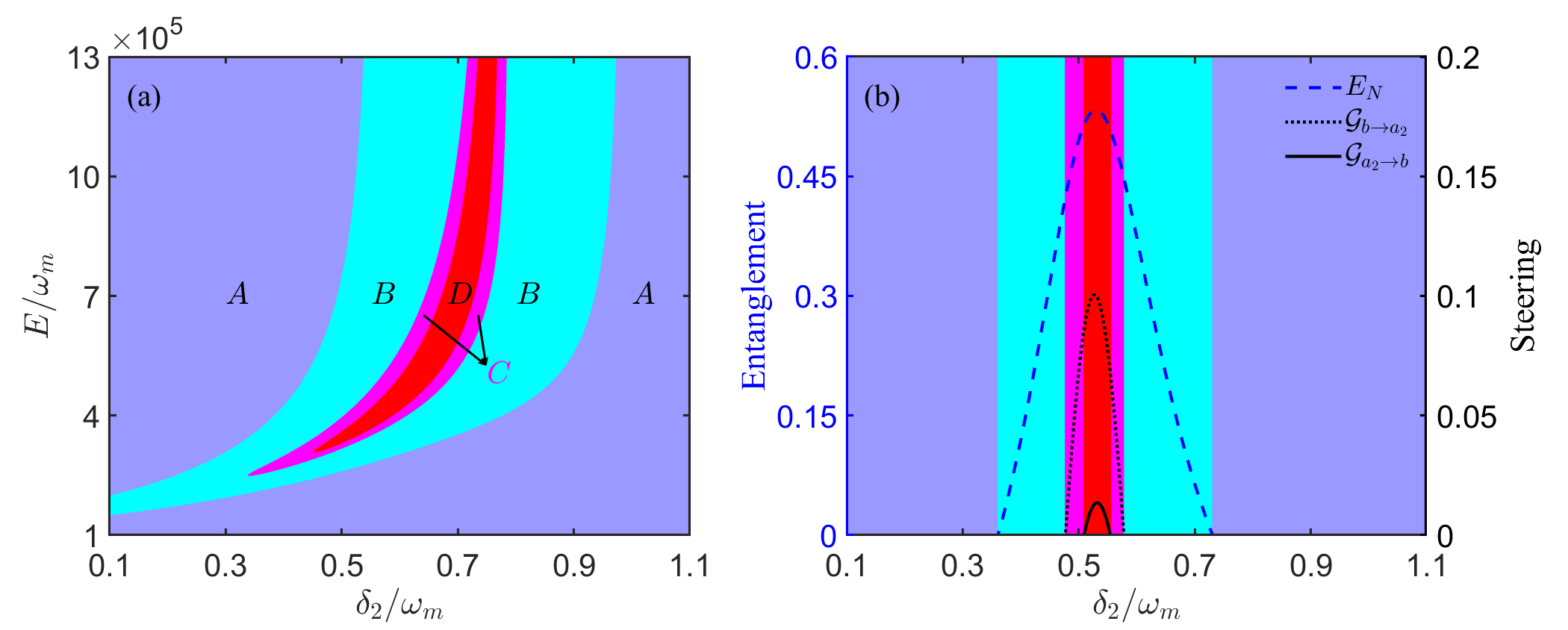}
\includegraphics[width=0.9\columnwidth]{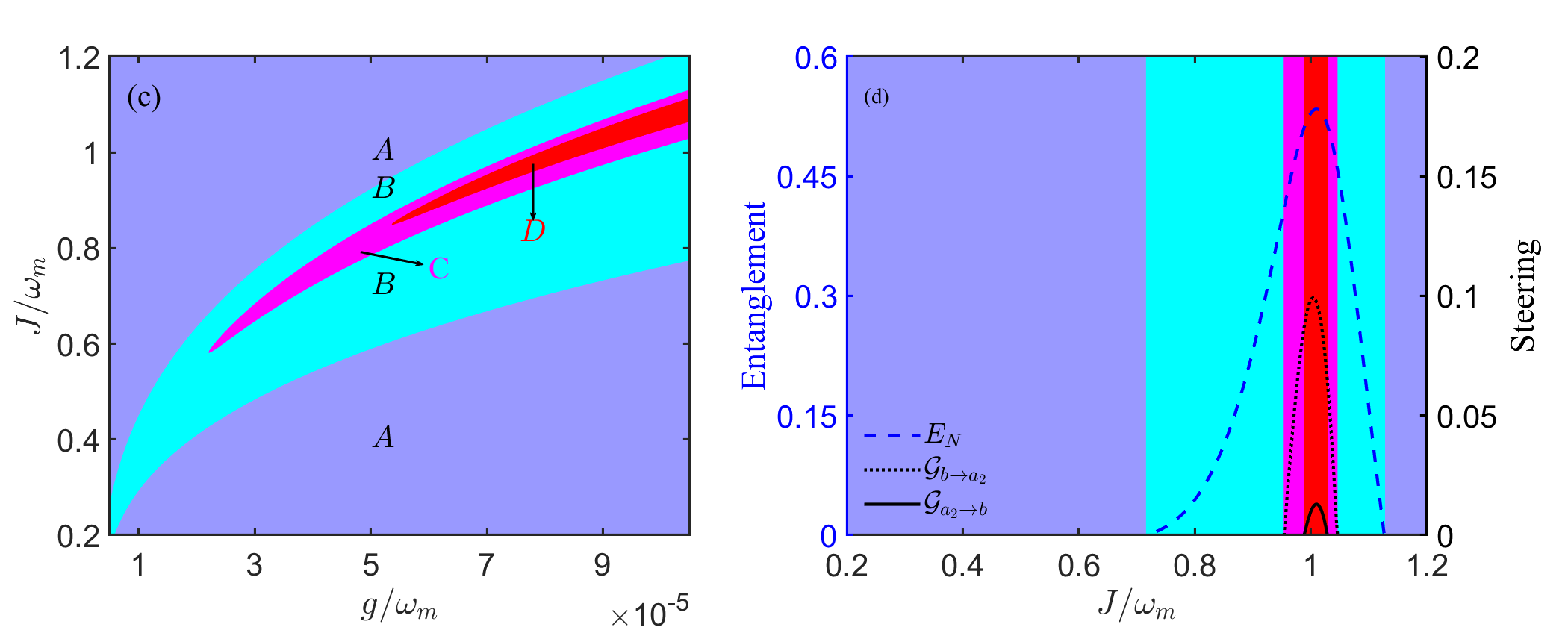}
\caption{The regimes of zero/nonzero entanglement, one-way steering and two-way steering. The lilac regimes are zero entanglement regions and marked by $A$; the cyan regimes are nonzero entanglement without steering regions and marked by $B$; the magenta regimes are one-way steering $\mathcal{G}_{b\rightarrow a_2}$ regions and marked by $C$; the red regimes are the two-way steering regions, and marked by $D$. In (b) and (d), the blue dashed line is the quantum entanglement; the black dotted line denotes the EPR steering $\mathcal{G}_{b\rightarrow a_2}$; and the black line shows the EPR steering $\mathcal{G}_{a_2\rightarrow b}$.}
\label{fig5}
\end{figure}

In panels (d)-(f), we show the variation in the quantum entanglement and EPR steering along with the single-photon coupling strength $g$ and resonator coupling strength $J$. The effective optomechanical coupling strength $G$ will be enhanced with increasing $g$, and the coefficients $\lambda$ and $\eta$ of the Bogoliubov mode $\beta$ will be enhanced synchronously. Based on the mechanism of entanglement generation, a larger $\lambda$ will lead to stronger entanglement. Moreover, a larger $\eta$ will reinforce the cooling effect on the Bogoliubov mode $\beta$, which will further strengthen the entanglement, as shown in panel (d), where the maximum value of entanglement increases monotonically with $g$. However, the entanglement cannot be increased indefinitely with increasing $g$. On the one hand, $\cosh\lambda$ and $\sinh\lambda$ will asymptotically equal with increasing $\lambda$, and the maximal entanglement will arrive asymptotically; on the other hand, the cooling effect on the Bogoliubov mode $\beta$ is limited due to the employment of the RWA approximation. When the interaction strength $\eta$ is strong enough, the counter rotating-wave term cannot be eliminated, and the corresponding Stokes effect will heat the Bogoliubov mode, which will counteract the cooling effect by the anti-Stokes effect and limit the maximal steady-state entanglement. In addition, it isn't easy to experimentally achieve a very strong single-photon coupled optomechanical system. A larger value of $J$ will decrease $G$ and $\eta$ and increase $\lambda$, which will lead to maximal entanglement for a modest value of $J$ due to the competition between $\lambda$ and $\eta$. The dynamics of EPR steering have a very similar physical mechanism, and its nonzero regions have distinct differences from quantum entanglement, which will be investigated in the following.

In Fig. \ref{fig5}, we plot regimes where the system has nonzero entanglement, one-way and two-way steering. The system parameters are chosen as those of maximal quantum entanglement in Fig. \ref{fig4}. The lilac regimes denote that there does not exist quantum entanglement, which are marked by regions $A$ in panels (a) and (c); the cyan regimes denote that there exists nonzero quantum entanglement, while the EPR steering is absent, which are marked by regions $B$; the magenta regimes (marked by regions $C$) denote that there only exists one-way steering $\mathcal{G}_{b\rightarrow a_2}$, i.e., the mechanical mode $\delta b$ can steer the resonator mode $\delta a_2$, while the resonator mode $\delta a_2$ cannot steer the mechanical mode $\delta b$; the red regimes, i.e., marked by regions $D$, present that there exists two-way steering, and both the modes $\delta a_2$ and $\delta b$ can steer each other. As a vivid exhibition, we plot the variation in quantum entanglement and EPR steering versus the detuning $\delta_2$ between the resonator $a_2$ and pumping laser in panel (b). The left axis denotes the quantum entanglement, which is expressed by the blue dashed line, while the EPR steering is described by the right axis with the black dotted line $\mathcal{G}_{b\rightarrow a_2}$ and the black line $\mathcal{G}_{a_2\rightarrow b}$. The variation in quantum entanglement and EPR steering versus the coupling strength $J$ is shown in panel (d), which can be realized by adjusting the gap between the resonators $a_1$ and $a_2$. From panels (b) and (d), one can find the transition from zero entanglement to strong entanglement and one-way steering to two-way steering through modulating the detuning $\delta_2$ or the coupling strength $J$, which is easy to implement experimentally.

\begin{figure}[b]
\centering
\includegraphics[width=0.9\columnwidth]{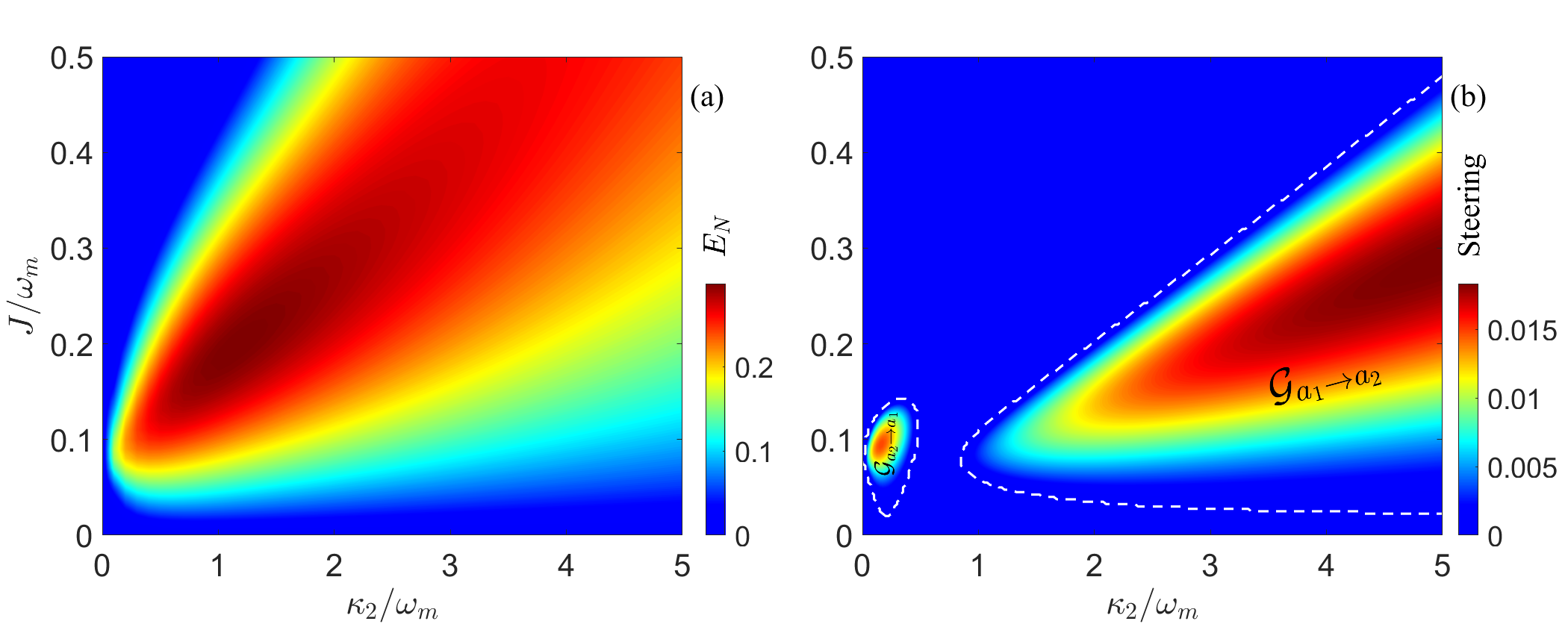}
\includegraphics[width=0.9\columnwidth]{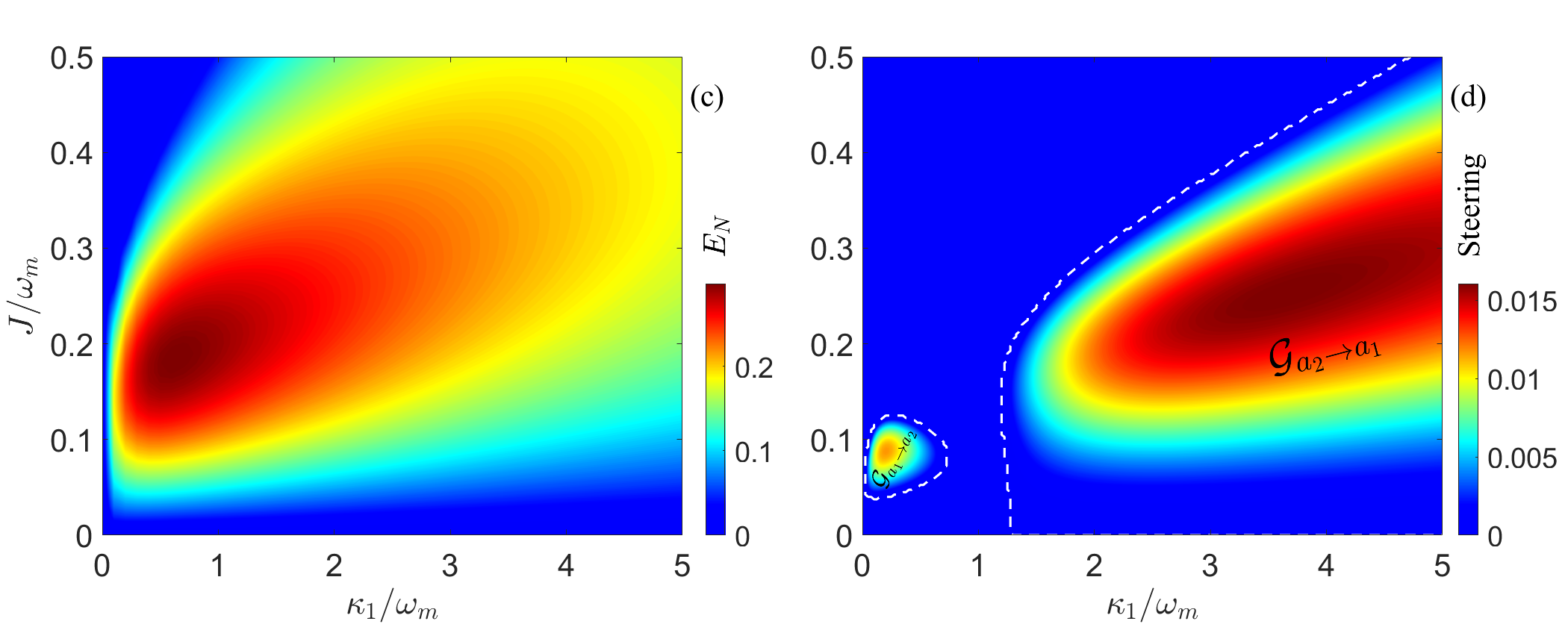}
\caption{The variation in quantum entanglement and EPR steering versus the coupling strength and resonator decay rate. The amplitude of the driving laser is $E/\omega_m=0.5\times10^4$, the detuning of the resonator $a_2$ and pumping laser is $\delta_2/\omega_m=0.8$, and the other parameters are the same as those in Fig. \ref{fig4}.}
\label{fig6}
\end{figure}

\section{Entanglement and steering of resonators $a_1$ and $a_2$}\label{sec4}
In this section, we will investigate the quantum nonlocality between the two resonators $a_1$ and $a_2$. In contrast to the quantum nonlocality of the resonator mode $\delta a_2$ and mechanical mode $\delta b$, there is no quantum entanglement when the amplitude $E$ of the driving laser is strong. According to Hamiltonian (\ref{eqHlin}), the mechanical mode $\delta b$ can be cooled by the optical mode $\delta a_1$ due to the beam-splitter interaction $G(\delta a_1\delta b^{\dagger}+\delta a_1^{\dagger}\delta b)$ under red-detuning driving, which will inhibit the quantum entanglement between resonators $a_1$ and $a_2$. The strength of effective interaction $G$ should be reduced to generate strong quantum entanglement. Due to $G=ga_{1s}$, there are two main approaches to decrease the value of $G$, i.e., weaken the driving amplitude $E$ or strengthen the squeezing parameter $r$. If the system's stability is considered, the value of $r$ should not be too large. When the red-detuing condition $\Delta_1'=\omega_m$ is employed, the effective coupling strength $G$ is mainly determined by the driving amplitude $E$, and the positive influence on the quantum entanglement can be realized by appropriately decreasing the value of $E$. In Fig. \ref{fig6}, we plot the quantum entanglement and EPR steering versus the coupling strength $J$ and resonators decay rate $\kappa_1$ and $\kappa_2$, respectively. Panels (a) and (c) show the quantum entanglement, and the variations in EPR steering are shown in panels (b) and (d). The detuning $\delta_2/\omega$ between the pumping laser and resonator $a_2$ is tuned to $0.8$, the driving amplitude is $E/\omega=0.5\times10^4$, and the other parameters are the same as those given in Fig. \ref{fig3}. One can find that the maximal quantum entanglement occurs in the vicinity of $\kappa_1\simeq\kappa_2$ and always in the regime in which $\kappa_2$ is slightly lager than $\kappa_1$. Intuitively, the quantum entanglement between two modes is symmetric, and the maximal quantum entanglement should occur where the two resonators have the same properties. It can be considered an additional decay path for the resonator $a_1$ with optomechanical interaction, such as $g/\omega_m=8.5\times10^{-5}$, so there needs to be a greater decay rate $\kappa_2$ to balance the ``additional decay rate'' of $\kappa_1$, which are shown in panels (a) and (c). Due to the direct interaction between the two resonators, the coupling strength $J$ should not be too large to exchange photons, which is in contrast to the optomechanical entanglement behavior in Fig. \ref{fig4}. Another striking feature of the entanglement between resonators $a_1$ and $a_2$ is the one-way EPR steering. The variation in EPR steering with decay rate $\kappa_2$ is shown in Fig. \ref{fig6}(b). The nonzero steering $\mathcal{G}_{a_2\rightarrow a_1}$ regime is in the lower left region, which means that mode $\delta a_2$ can steer mode $\delta a_1$; while the nonzero steering $\mathcal{G}_{a_1\rightarrow a_2}$ regime is in the right region, which denotes that mode $\delta a_1$ can steer mode $\delta a_2$. In the stable regime, there does not exist two-way EPR steering, which is markedly different from the steering between the resonator $a_2$ and mechanical oscillator $b$. By controlling the decay rate, we can manipulate and utilize unidirectional EPR steering to implement a desired quantum process. Similarly, the EPR steering versus the decay rate $\kappa_1$ is shown in panel (d).

\begin{figure}[h]
\centering
\includegraphics[width=0.5\columnwidth]{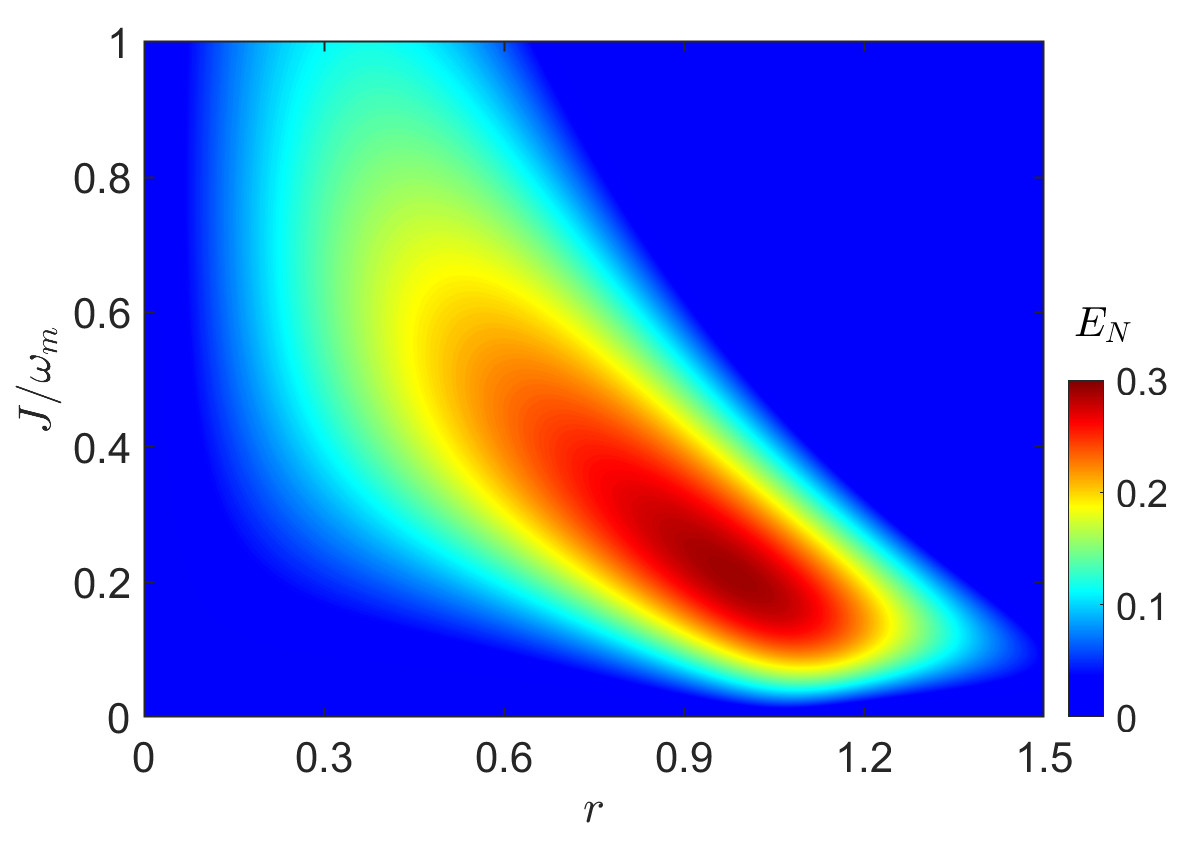}
\caption{The variation in quantum entanglement along with the coupling strength and squeezing parameter. The system parameters are the same as those in Fig. \ref{fig6}.}
\label{fig7}
\end{figure}

The quantum entanglement between resonators $a_1$ and $a_2$ versus the coupling strength $J$ and squeezing parameter $r$ are shown in Fig. \ref{fig7}. One can find that the coupling between resonators $a_1$ and $a_2$ and the squeezing of resonator $a_2$ induced by the pumping laser is necessary to generate quantum entanglement. Even though the squeezing parameter $r$ can strengthen the effective coupling strength $J^s$ between the modes $\delta a_1$ and $\delta a_2$, the large $J^s$ harms entanglement, which is similar to the analysis in Fig. \ref{fig6}. The strength of the original coupling $J$ should also be moderate, and there is a similar detrimental effect to entanglement with large $J$. Therefore, the key to generating strong entanglement is optimizing the squeezing parameter $r$ based on the system parameters rather than simply raising the value of $r$.

\section{Conclusions}
Quantum entanglement and EPR steering are important characteristics that distinguish  quantum systems from the classical world and significant physical resources in modern quantum technology, and how to generate quantum entanglement and EPR steering is at the heart of related research. This paper proposes a theoretical proposal to manipulate optomechanical entanglement and EPR steering induced by quantum squeezing in a coupled WGM optomechanical system. By pumping the $\chi^{(2)}$-nonlinear resonator with a phase matching condition, strong entanglement and steering between the mechanical oscillator and the indirect interacted squeezed resonator can be generated, where the squeezing of the $\chi^{(2)}$-nonlinear resonator plays a vital role in this process. By adjusting the system parameters, such as the detuning of the pumping $\chi^{(2)}$-nonlinear resonator or the coupling strength between the two resonators, the transitions from zero entanglement to strong entanglement and one-way steering to two-way steering can be realized in the stable regime. Meanwhile, the photon-photon entanglement of the coupled resonators can be obtained by weakening the driving laser's amplitude, and nonreciprocal one-way steering can be realized by changing the decay rates. This project paves a flexible, convenient, and innovative avenue to utilize optomechanical or photon-photon entanglement in quantum technology since there is no extraordinary squeezed field in this project, and entanglement and steering can be realized by controlling the system parameters. Because multi-body entanglement has richer properties than two-body entanglement and may have greater advantages in quantum processes, the multi-body entanglement among the three modes in this project and the role of squeezing in it are worthy of systematic investigation in the following works.

\begin{backmatter}
\bmsection{Funding}
National Key Research and Development Program of China (2021YFA1402002), National Natural Science Foundation of China (U21A20433, 12175029, and 12204440), Fundamental Research Program of Shanxi Province (20210302123063, and 202103021223184).

\bmsection{Disclosures}
The authors declare no conflicts of interest.

\bmsection{Data Availability}
The data that support the findings of this study are available upon reasonable request from the authors.
\end{backmatter}

\newcommand{\doi}[2]{\href{https://doi.org/#1}{\color{blue}#2}}

\end{document}